\documentclass[aps,pre,twocolumn,floatfix,showpacs]{revtex4}

\usepackage{graphicx}
\bibstyle{apsrev.bst}

\begin{document}
\title{Thermodynamics of RNA/DNA hybridization in high density
oligonucleotide microarrays}
\author{Enrico Carlon and Thomas Heim}
\affiliation{Interdisciplinary Research Institute c/o IEMN, Cit\'e
Scientifique BP 60069, F-59652 Villeneuve d'Ascq, France}

\begin{abstract}
We analyze a series of publicly available controlled experiments (Latin
square) on Affymetrix high density oligonucleotide microarrays using
a simple physical model of the hybridization process.  We plot for
each gene the signal intensity versus the hybridization free energy
of RNA/DNA duplexes in solution, for perfect matching and mismatching
probes. Both values tend to align on a single master curve in good
agreement with Langmuir adsorption theory, provided one takes into
account the decrease of the effective target concentration due to
target-target hybridization in solution. We give an example of a
deviation from the expected thermodynamical behavior for the probe set
1091\_at due to annotation problems, i.e. the surface-bound probe is
not the exact complement of the target RNA sequence, because of errors
present in public databases at the time when the array was designed.
We show that the parametrization of the experimental data with RNA/DNA
free energy improves the quality of the fits and enhances the stability
of the fitting parameters compared to previous studies.
\end{abstract}

\pacs{87.15.-v,82.39.Pj}

\date{\today}

\maketitle
\newcommand{\ul}{\underline}
\newcommand{\bc}{\begin{center}}
\newcommand{\ec}{\end{center}}
\newcommand{\be}{\begin{equation}}
\newcommand{\ee}{\end{equation}}
\newcommand{\ba}{\begin{array}}
\newcommand{\ea}{\end{array}}
\newcommand{\beqn}{\begin{eqnarray}}
\newcommand{\eeqn}{\end{eqnarray}}

\section{Introduction}
\label{sec:intro}

\begin{figure}[b]
\includegraphics[width=6cm]{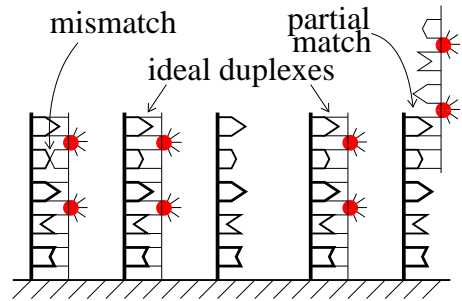}
\caption{DNA microarrays are based on the hybridization of surface-bound
DNA probes (thick) with target sequences in solution carrying fluorescent
labels (thin).  Besides perfect matching probe-target pairs forming
ideal duplexes, partial hybridizations, or mismatches are possible,
although they are expected to be thermodynamically less stable.}
\label{FIG01}
\end{figure}

DNA microarrays \cite{brow99,lips99_sh} are devices capable of measuring
the gene expression levels on a genome-wide scale and are based on the
{\it hybridization} between surface-bound DNA sequences (the probes) and
DNA, or RNA, sequences in solution (the targets). While the specificity
of the interaction between complementary base pairs A--T and C--G
suggests that the hybridization of a single stranded DNA target with its
perfect matching probe would be dominant, often strong non-complementary
hybridization effects are observed (see Figure \ref{FIG01}). As the
targets are fluorescently labeled, the amount of hybridized DNA from
each probe can be determined from optical measurements. The presence
of strong cross-hybridizations is one of the reasons why one cannot
interpret the fluorescent light intensities as direct measures of the
gene expression levels.

One of the most popular platform for DNA microarrays is provided by
Affymetrix \cite{lips99_sh}, which produces high-density oligonucleotide
arrays. In Affymetrix chips short single stranded DNA sequences ($25$
nucleotides) are grown {\it in situ} using photolitographic techniques.
As a single probe of just $25$ nucleotides may not provide enough
specificity for a reliable measurement of the gene expression level, a
set of 10-16 probes (the probe set) complementary to different regions
of the same target sequence are present in the chip.  For each perfect
matching (PM) probe there is a sequence differing by a single nucleotide.
These are referred to as mismatching (MM) probes and are used to quantify
the effects of cross-hybridization \cite{lips99_sh}.

Most of the available software packages for the calculation
of gene expression levels from the fluorescence intensities
rely on algorithms of purely statistical or empirical nature
\cite{liwo01,bols03_sh}. In the past two years, however, several
algorithms based on physical properties of the hybridization process
were proposed \cite{zhan03,heks03_sh,naef03,held03,deut04}.  The basic
idea behind the latter approach is that the intensities are linked to
the hybridization free energies for the formation of the probe-target
duplexes. For instance, for equal target concentration in solution,
binding to CG-rich probes will provide a stronger signal compared to
CG-poor probes (CG nucleotides are more strongly bound than AT pairs
\cite{bloo00}).

\begin{figure}[t]
\includegraphics[width=6cm]{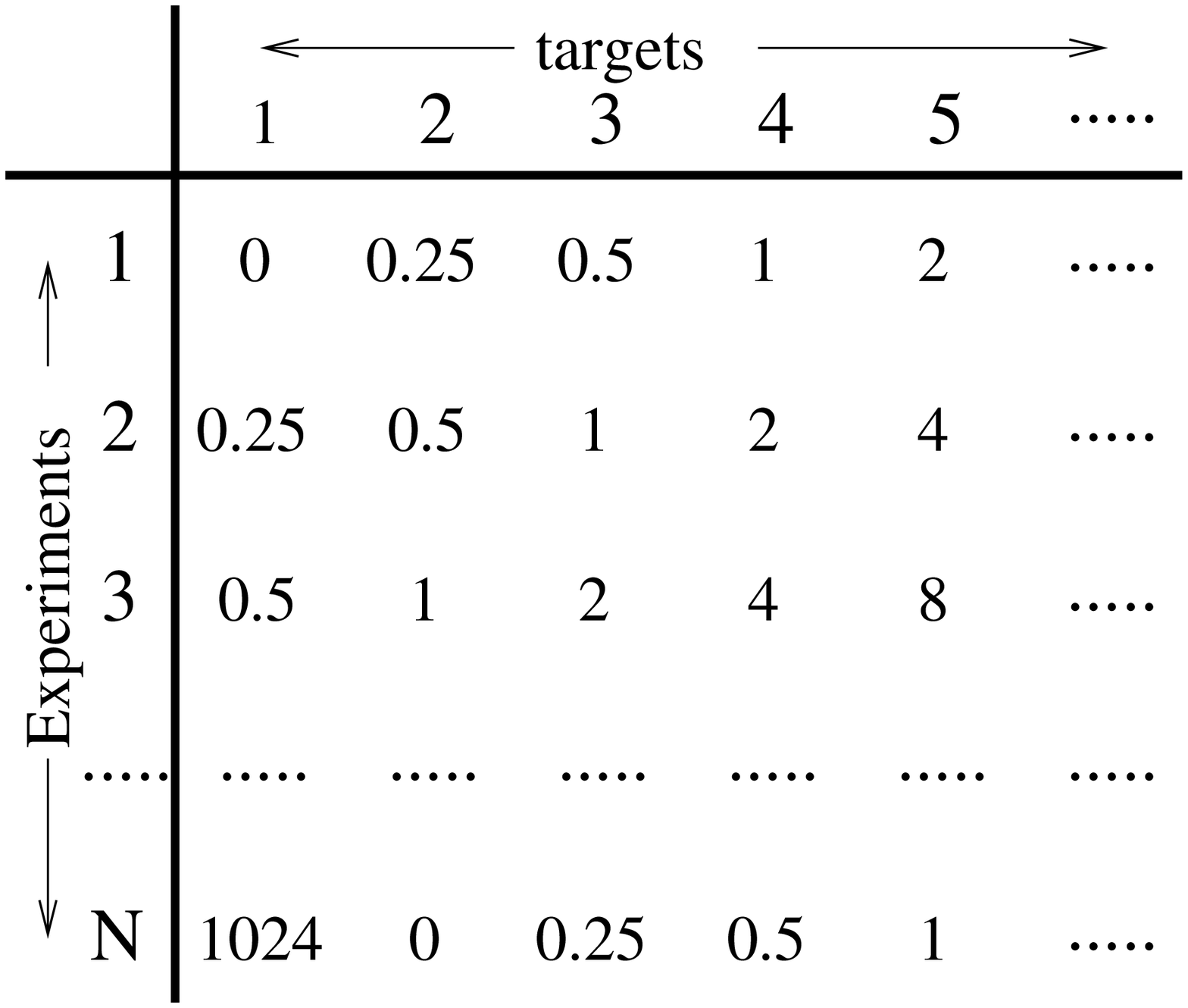}
\caption{In the Latin square experiments some selected target sequences
are added at known concentrations following the scheme indicated in
the figure.  Affymetrix considered 14 different concentrations ranging
from $0$ to $1024$ pM (picomolars) and varying by a factor 2. In the
Experiment 1, for instance, the RNA target 1 is absent from the solution
($0$ pM), the target 2 is present at a concentration of $0.25$ pM \ldots
In this scheme all possible 14 concentrations for each of the 14 target
sequences are explored in 14 different experiments.}
\label{latin_square}
\end{figure}

In this paper we investigate a set of controlled experiments known as
Latin square experiments and performed by Affymetrix in the human
HGU95a chipset. In these experiments some target sequences are
added at controlled (``spike-in") concentrations on a background
reference solution.  The target concentrations range from $0$ to
$1024$ pM increasing as a power of $2$ and following the scheme
depicted in Fig. \ref{latin_square}, covering all concentrations of
biological interest. Note that in the table experiment vs. targets of
Fig. \ref{latin_square} equal concentrations are found along the lower
left-upper right diagonals, following thus a pattern known as Latin
square.  The data, which are publicly available from the Affymetrix web
site \cite{affy}, are important references for testing new algorithms
that calculate gene expression levels from ``raw" microarrays data. Not
surprisingly, due to their central importance, the Latin square
data have been analyzed by several groups through various approaches
\cite{liwo01,bols03_sh,zhan03,heks03_sh,naef03,held03,deut04,bind04,burd04}.
Our analysis is based on a simple physical model of target-probe
hybridization. Although the modeling of microarray data with the physics
of hybridization has been followed by other groups in the past couple of
years \cite{zhan03,naef03,held03}, our approach differs from what has been
done so far in the following ways: 1) We use the free energy parameters of
formation of RNA/DNA duplexes in solution, and not the DNA/DNA parameters
as in \cite{held03}. 2) We include the analysis of mismatches. 3) We
include the effect of target-target hybridization in solution.

The latter effect turns out to be an essential feature of our approach:
When target-target hybridization is neglected the fit of the experimental
data is very poor for half of the 14 spike-in genes. On the contrary,
when hybridization in solution is included we obtain good fits of
the experimental data with a simple theory containing four fitting
parameters.  The ultimate test of the validity of our approach is
through the analysis of scaling collapses: when plotted as a function
of an appropriate rescaled thermodynamic variable, which depends on an
effective temperature, on the hybridization free energies and on the
target concentration, the Latin square data for different experiments
tends to collapse into a single master curve. Although the noise level
can still be large, significant deviations from this master curve are
very rare. As we shall see, the deviations from the expected isotherm
can be understood in several cases.

This paper is organized as follows: in Section \ref{sec:RNA_DNA}
we discuss the thermodynamic parameters used in this paper. In
Sec. \ref{sec:Affy} we present the analysis of the Latin square data and
the model used. In Section \ref{sec:deviations} we present few examples
of probes deviating from the expected behavior and discuss the origin of
these deviation. Section \ref{sec:Conclusion} concludes the paper with
an overview of the results, open issues and a discussion of related works.

\begin{table}[h]
\label{tableI} 
\caption{The stacking free energy parameters $\Delta G_{\rm
37}$ for RNA/DNA hybrids measured in solution at a salt concentration
$1$ M NaCl and $T=37^\circ$ C \cite{sugi95_sh}. The upper strand is RNA
(with orientation $5'$-$3'$) and lower strand DNA (orientation $3'$-$5'$).
Between parenthesis we give the DNA/DNA parameters.}
\vspace{5mm}
\begin{tabular}{@{}cc|cc}
\hline
Seq.&$-\Delta G_{\rm 37}$ (kcal/mol) &Seq.& $-\Delta G_{\rm 37}$ (kcal/mol)\\
\hline
&&&\\
${\rm rAA} \atop {\rm dTT}$ & 1.0 (1.00)&
${\rm rAC} \atop {\rm dTG}$ & 2.1 (1.44)\\
&&&\\
${\rm rAG} \atop {\rm dTC}$ & 1.8 (1.28)&
${\rm rAU} \atop {\rm dTA}$ & 0.9 (0.88)\\
&&&\\
${\rm rCA} \atop {\rm dGT}$ & 0.9 (1.45)&
${\rm rCC} \atop {\rm dGG}$ & 2.1 (1.84)\\
&&&\\
${\rm rCG} \atop {\rm dGC}$ & 1.7 (2.17)&
${\rm rCU} \atop {\rm dGA}$ & 0.9 (1.28)\\
&&&\\
${\rm rGA} \atop {\rm dCT}$ & 1.3 (1.30)&
${\rm rGC} \atop {\rm dCG}$ & 2.7 (2.24)\\
&&&\\
${\rm rGG} \atop {\rm dCC}$ & 2.9 (1.84)&
${\rm rGU} \atop {\rm dCA}$ & 1.1 (1.44)\\
&&&\\
${\rm rUA} \atop {\rm dAT}$ & 0.6 (0.58)&
${\rm rUC} \atop {\rm dAG}$ & 1.5 (1.30)\\
&&&\\
${\rm rUG} \atop {\rm dAC}$ & 1.6 (1.45)&
${\rm rUU} \atop {\rm dAA}$ & 0.2 (1.00)\\
&&&\\
\hline
\end{tabular}
\end{table}

\section{Thermodynamics of RNA/DNA hybrids}
\label{sec:RNA_DNA}

The thermodynamics of duplex formation of nucleic acids in solution is
well described by the nearest neighbor model according to which the free
energy difference between a duplex and two separated strands is given
by the sum of the local terms which keep into account hydrogen bonding
and base stacking \cite{bloo00}. In melting experiments in solution one
usually determines $\Delta H$ and $\Delta S$ the enthalpy and entropy
differences between a duplex and two separate strands, from which the
free energy difference $\Delta G = \Delta H - T \Delta S$ is obtained. The
Table \ref{tableI} gives the free energy parameters at $1$ M of NaCl and
at $T=37^\circ$ C (data taken from Ref. \cite{sugi95_sh}). The calculation
of $\Delta G$ for a given sequence is obtained by summing up the data on
Table \ref{tableI} and adding to that a contribution of helix initiation
$\Delta G^{\rm init.}_{37} = 3.1$ kcal/mol \cite{sugi95_sh}.

The thermodynamic parameters for RNA/DNA hybrids containing a single
mismatch have recently been determined \cite{sugi00}. The simple nearest
neighbor model with stacking free energy parameters is no longer accurate
for mismatches. For RNA/DNA single mismatches it has been found that the
trinucleotide model, in which distinct free energies are associated to
the triplet formed by the mismatch and the two neighboring nucleotides,
fits the experimental data reasonably well \cite{sugi00}.  As a MM
probe in Affymetrix chips is realized by interchanging C with G and
A with T in the middle nucleotide of a PM probe, there are four types
of mismatches rGdG, rCdC, rUdT and rAdA. Taking into account the four
possible combinations of neighboring nucleotides there are thus in total
64 different mismatches that should be considered. The free energy of
only part of these 64 triplets can be found in the present literature
\cite{sugi00}.  The full list of mismatch free energies used in this
paper is given in the Appendix \ref{sec:appendix}.

\section{Latin square data}
\label{sec:Affy}

Usually, for the intensities $I$ measured in the Affymetrix experiment
one distinguishes the two contributions from non-specific ($N$) and
specific ($S$) hybridizations \cite{liwo01}. We follow the same idea
here and write:
\be
I(c, \Delta G ) = N +  S(c, \Delta G) + \varepsilon
\label{specific_non-specific}
\ee
where $\varepsilon$ denotes some experimental noise.  Here $I$ is the
intensity from the probe whose complementary RNA target is in solution
at a concentration $c$ (known for the ``spike-in" genes) and $\Delta G$
the hybridization free energy. Note that we did not make any distinction
between PM and MM probes, as we assume that their specific binding will
depend only on $c$ and $\Delta G$. The non-specific hybridization $N$
depends on the total RNA concentration in solution and possibly on
other free energy parameters describing the partial matching with all
RNA in solution. However, the precise form of $N$ is not relevant for
the analysis performed in this paper, as we focus here on
\be
\Delta I \equiv I(c) - I(0) \approx S(c, \Delta G)
\label{low_spike}
\ee
The background subtraction is only possible in the Latin square
set as there is always a reference measurement at zero spike-in
concentration. The problem of calculating the ``background" ($N$ in
Eq. (\ref{specific_non-specific})) from first principles approaches will
be addressed elsewhere.

As in Ref. \cite{held03}, we model the specific hybridization as a
two-state process where the target is either unbound in solution or
fully hybridized to the probe forming a 25 nucleotides double helix
at the surface, with one mismatch at position 13 for the MM probes.
The Langmuir model predicts that:
\be
S (c, \Delta G) = \frac{A c e^{-\beta \Delta G}}{1 + c e^{-\beta \Delta G}}
\label{langmuir}
\ee
here $\beta=1/RT$, where $T$ is the temperature and $R=1.99$ ${\rm
cal/mol\cdot K}$ is the gas constant.  The parameter $A$ sets the scale
of the intensity and corresponds to the saturation value in the limit
where $c \gg e^{\beta \Delta G}$, i.e. where the concentration is high
or the binding is strong.

\begin{figure}[h]
\includegraphics[width=8cm]{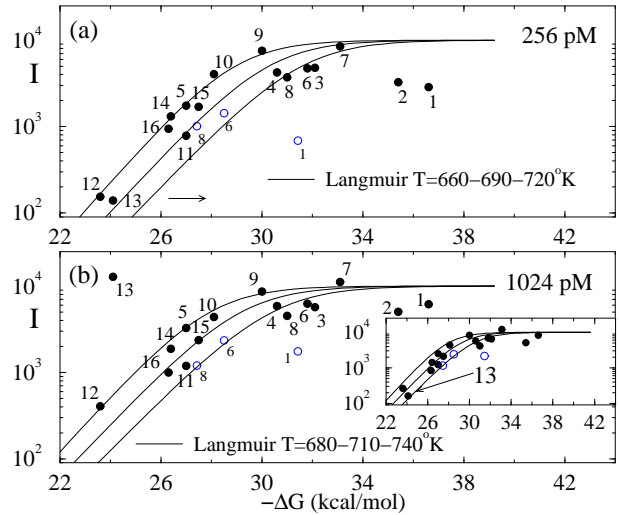}
\caption{Signal intensities versus the bulk hybridization free energy
for the probe set 1708\_at, ``spiked-in" at concentration $256$ pM (a)
and $1024$ pM (b). The data are obtained from the Affymetrix experiments
1521m99hpp\_av06 and 1521a99hpp\_av06, respectively.  Filled and empty
symbols refer to PM and MM probe sets. The solid lines are curves
from Eq. (\ref{langmuir}) where $c$ is given in the experiment,
$A=10^4$ and for three values of the temperature (the arrow in (a)
indicates the direction of increasing temperature).  The inset shows
a plot for a replicate of the $1024$ pM experiment taken from the file
1532a99hpp\_av04. Notice that in the latter the intensity of the PM probe
13 (indicated by the arrow) agrees very well with the Langmuir isotherm.}
\label{FIG03}
\end{figure}

\subsection{High ``spike-in" concentrations}

It is convenient to analyze first the limit of high ``spike-in"
concentrations, which we find to correspond to $c \geq 256$ pM. At such
high concentrations typically $I(c) \gg I(0)$ thus the contribution of
the background signal can be safely ignored. It is therefore equivalent
to plot $I(c)$ or $\Delta I(c)$, as defined in Eq. (\ref{low_spike}).

Figure \ref{FIG03} shows a plot of $I$ vs. $\Delta G$ for the probe set
1708\_at for the concentrations of $256$ (a) and $1024$ pM (b). Both PM
and MM probes (filled and empty symbols) are shown. The numbers label
the probes, following the notation chosen by Affymetrix.  For the MM
probes we could calculate $3$ out of $16$ free energies, using the data
given in Table \ref{tableII}. Although fluctuations are quite strong,
the intensities shown in Figure \ref{FIG03} tend to align to a single
master curve both for PM and MM probes.

The solid lines in Figure \ref{FIG03} are plots of the Langmuir isotherm
given in Eq. (\ref{langmuir}). Assuming that the probe density is
roughly constant for the whole array, we expect that the value of the
saturation amplitude $A$ is the same for all probes. We analyzed the
histograms of intensities for the whole set of Latin square experiments
and found that the probability of finding an intensity $I$ drops sharply
beyond $I_{\rm max} \approx 10^4$, therefore we fix in the whole paper
$A=10^4$. As the concentration $c$ is known, the only free parameter is
the temperature $T$.  We find that best fits of the Langmuir isotherm
with the experimental data are obtained for a temperature $T=700 \pm 30
^\circ K$, roughly twice as large as the temperature in the Affymetrix
experiment, which is of $T = 45^\circ C \approx 320 ^\circ K$.

\begin{figure}[t]
\includegraphics[width=8cm]{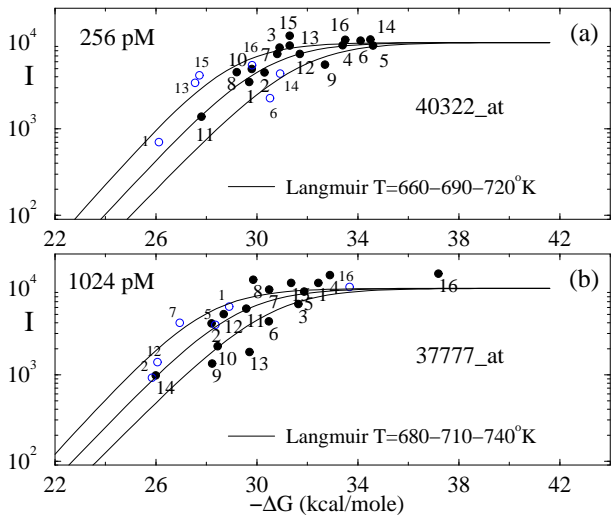}
\caption{As in Figure \ref{FIG03} for the probe set 40322\_at. The data are
obtained from Affymetrix experiments given in the files 1521b99hpp\_av06
(a) and 1521d99hpp\_av06 (b).}
\label{FIG04}
\end{figure}

The ``discrepancy" between fitted and experimental temperatures deserves
some discussion. We have estimated $\Delta G$ from a two state model
summing up over the stacking parameters of Table \ref{tableI}. In reality
the binding of a target with a PM or MM probe can also involve fewer
nucleotides. Moreover, the photolitographic process is not perfect and
the surface bound probes have varying lengths (see Ref. \cite{form98}).
These remarks indicate that the binding free energies are lower than those
we have estimated on the basis of a simple two state process assuming
that all probes have a fixed length of 25 nucleotides.  However, we
note that in plots of intensities versus $\Delta G$, the hybridization
free energies calculated from Table \ref{tableI}, the experimental data
tend to align along a single master curve, as shown in Fig. \ref{FIG03}
and in the rest of the paper. This suggests that $\Delta G$ is a good
thermodynamic variable to parametrize the experimental data. The fact
that the data follow a Langmuir isotherm suggests also that differences
with the true hybridization free energy in the array can be reabsorbed
in a rescaling of the temperature, as $\Delta G$ enters in the analysis
through a Boltzmann weight $\exp (-\beta \Delta G)$. An ``effective"
temperature of about $700^\circ$ K implies that on average $\Delta
G_{\rm array} \approx \Delta G_{\rm sol}/2$. Being an ``effective"
parameter  $T$ should not be compared directly to the experimental value.
More important, for the purposes of this work, is the stability of
$T$ as a fitting parameter: our analysis indicates that $T=700^\circ$
K fits rather well the experimental data for different probe sets and
spike-in concentrations.

A rather high effective temperature ($T = 2100^\circ$ K) was found in
the fit of the Latin square data of Ref. \cite{held03}. The difference
between our estimate and that of Ref. \cite{held03} is due to a different
free energy parametrization (we use the more appropriate RNA/DNA values)
and a different fitting procedure. Here we focus on fits of Langmuir
isotherms as function of $\Delta G$, rather than as function of the
concentration as done in \cite{held03}. These issues are discussed in
Appendix \ref{sec:app_held}.

In Figure \ref{FIG03} one notices the presence of few ``outliers",
i.e. those probes whose intensities strongly deviate from the Langmuir
isotherm, for instance as the probe 13 in Figure \ref{FIG03}(b).
The inset of Figure \ref{FIG03} shows a replicate of the experiment at a
concentration of $1024$ pM. In that case the intensity of probe 13 is in
agreement with the Langmuir isotherm. The intensities from the probes 1
and 2 instead deviate systematically from the Langmuir isotherm in all
replicates of the 256 pM and 512 pM experiments. The origin of these
deviations is discussed below.

Figure \ref{FIG04} shows the intensities for the probe set 40322\_at
for ``spike-in" concentrations of $256$ pM (a) and for the probe set
37777\_at at a concentration of $1024$ pM (b). Again the trend of the
PM and MM data is to align into a single master curve fitting quite well
Eq. (\ref{langmuir}), when the same effective temperatures as in Figure
\ref{FIG03} are used. Similar behavior is found for the other spike-in
genes \cite{affy_online}.

\subsection{General case}

In order to test the global functional form of the Langmuir isotherm
we turn now to the analysis of the full range of concentrations. From
Eq. (\ref{langmuir}) we expect that the experimental data should
``collapse" into a simple master curve when plotted as a function of the
scaled variable $x=c \ \exp(-\beta \Delta G)$. A preliminary analysis
at various temperatures at around $T=700^\circ$ K shows that the best
fits are obtained for an effective temperature $T = 680^\circ$ K, which
we fix now once for all.

\begin{figure}[t]
\includegraphics[width=8cm]{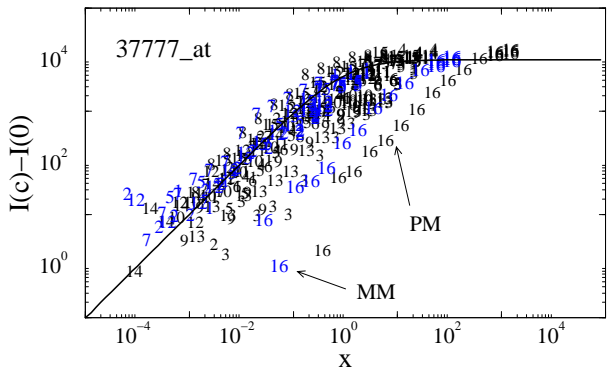}
\caption{Intensities for the Latin square experiment (set 1521
\cite{affy}) for the probe set 37777\_at plotted as function of the
rescaled variable $x= c \exp (\beta \Delta G)$.  The probe numbers for 
both PM (smaller characters) and MM (bigger characters) are given.}
\label{FIG09}
\end{figure}

\begin{figure}[t]
\includegraphics[width=8cm]{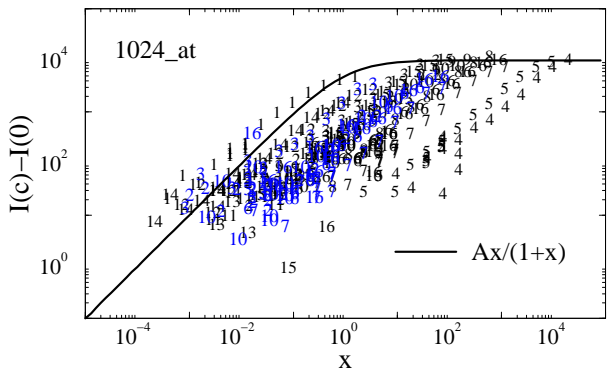}
\caption{Plot of $\Delta I$ vs. $x=c \exp(-\beta \Delta G)$ for the
probe set 1024\_at.}
\label{FIG10}
\end{figure}

Figure \ref{FIG09} shows a plot of $\Delta I$, as defined in
Eq. (\ref{low_spike}), vs. $x$ for the probe set 37777\_at.  Note that
the large majority of probes follow indeed the Langmuir isotherm
which takes the form $Ax/(1+x)$, and which is shown as a solid line
in Figure \ref{FIG09}.  Only the intensities of the probe 16 deviate
substantially from it.  Quite interestingly, probe 16 still follows
a Langmuir isotherm shifted along the horizontal axis.  This shift
is equivalent to a probe-dependent rescaling of the variable $x$. One
can thus collapse all the data onto the curve $Ax'/(1+x')$ by plotting
$\Delta I$ as a function of
\be 
x' = \alpha_k \ c \ e^{-\beta \Delta G} , 
\label{x_k} 
\ee 
with $\alpha_k$ probe dependent. For instance, for the probe set 37777\_at
one could take $\alpha_k  \approx 1$ for all probes except for probe 16
for which $\alpha_{16} \approx 10^{-3}$. While in Figure \ref{FIG09}
only one probe deviates sensibly from the Langmuir isotherm, in other
cases the disagreement involves the majority of the probes. An example
is given in Figure \ref{FIG10}, which shows a plot of $\Delta I$ vs. $x$
for the probe set 1024\_at.  We note that the shift along the $x$-axis is
predominantly to the right side of the Langmuir isotherm, corresponding
to a rescaling parameter $\alpha_k < 1$. An analysis of all the 14
``spike-in" genes shows that half of them are quite well-behaving in
the sense that most of the data on a plot of $\Delta I$ vs. $x$ align
along the Langmuir isotherm, as in Figure \ref{FIG09}. The remaining half
resembles more the example of Figure \ref{FIG10} (all figures are shown
in \cite{affy_online}).  A closer inspection to these defective probes
shows that most of them have a rather high hybridization free energy,
typically larger than $30-35$ kcal/mol. 

\begin{figure}[t]
\includegraphics[height=4cm]{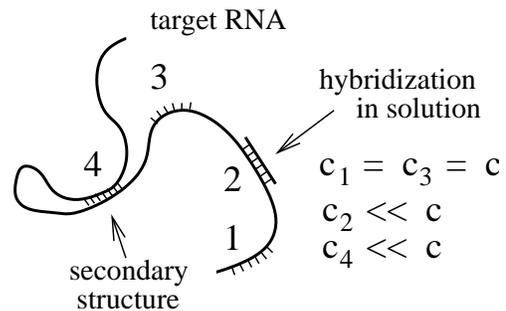}
\caption{Example of reduction of the effective concentration in solution
of the target sequences due to hybridization with other RNA fragments in
solution (2) or due to secondary structure formation (4). For probes 2 and
4 the target sequences available for hybridization is decrease.  We model
this effects with a simple function $\alpha$ decreasing exponentially
in the RNA/RNA hybridization free energy [Eq. (\ref{alpha_k})].}
\label{FIG11}
\end{figure}

A rescaling factor $\alpha < 1$ can also be interpreted as a lowering
of the target concentration in solution $c'= \alpha c$. The most
plausible explanation of this reduction of the concentration is the
target-target hybridization in solution, or RNA secondary structure
formation, as schematically illustrated in Figure \ref{FIG11}. The figure
shows an example four 25 nucleotides long regions of the target RNA,
which are complementary to probes 1, 2, 3 and 4 of some probe set. The
regions richer in CG, which have therefore higher hybridization free
energy (in the example 2 and 4), tend to form stable duplexes with other
RNA fragments or to form some secondary structure.  Once hybridization in
solution has occurred the amount of target RNA available for hybridization
to the probe sequences is reduced.

\begin{figure}[t]
\includegraphics[width=8cm]{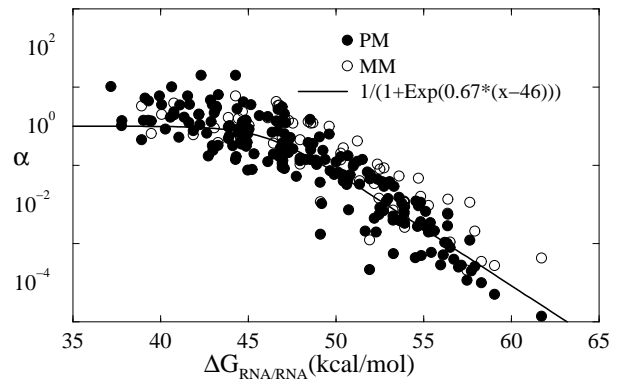}
\caption{Plot of the rescaling factor $\alpha$ needed to shift the data
points to the Langmuir isotherm $Ax/(1+x)$ as function of the RNA/RNA
hybridization free energies for each probe. The solid line is the fit
to the Eq. (\ref{alpha_k}) as expected from a simple model of bulk
hybridization.  The circles and diamonds emphasize the data from the
probe sets 408\_at and 36889\_at which contain some defective probes.
All these probes tend to deviate more strongly from the average behavior.}
\label{FIG12}
\end{figure}

\begin{figure*}[t]
\includegraphics[height=5cm]{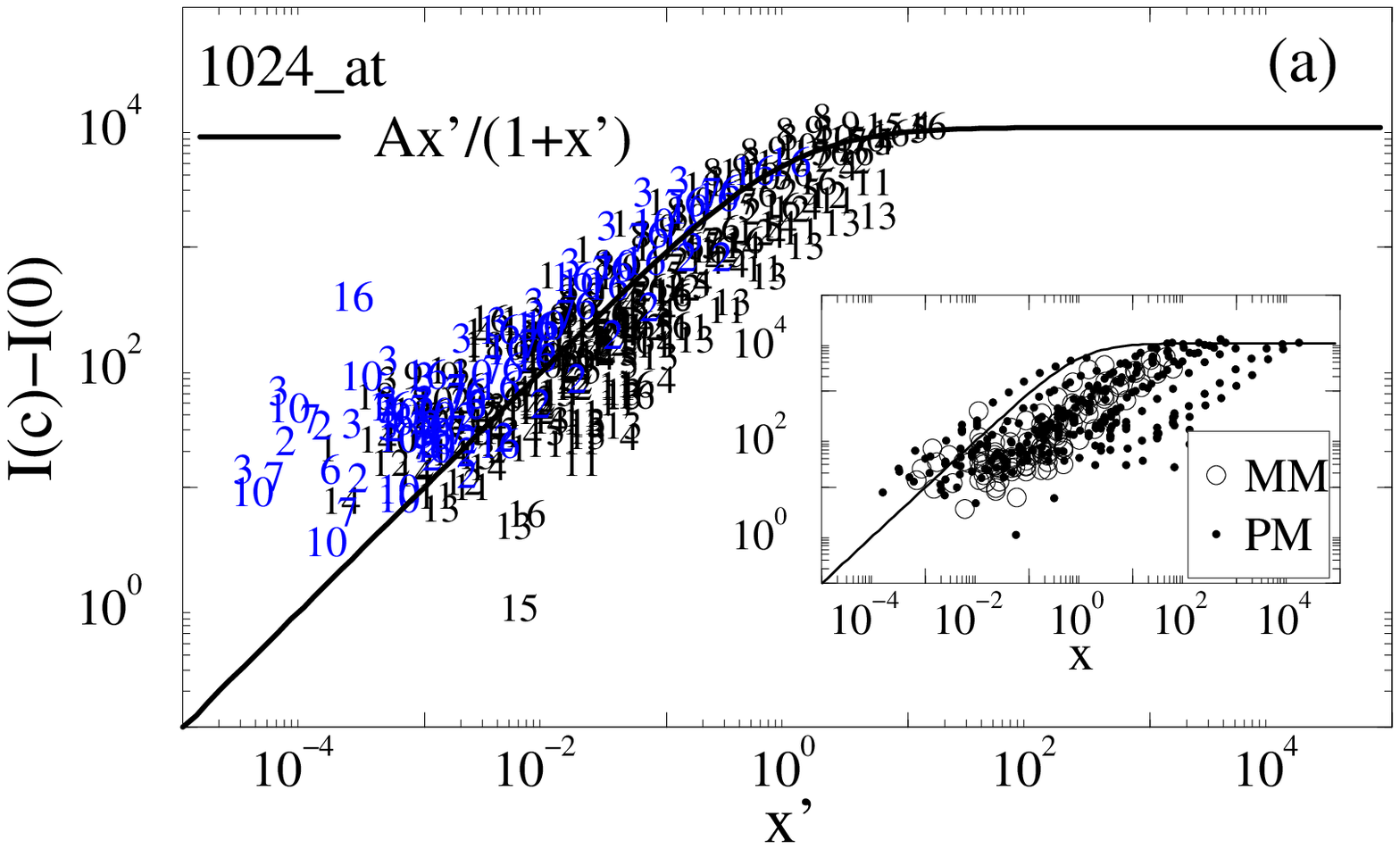}
\ \ \ \
\includegraphics[height=5cm]{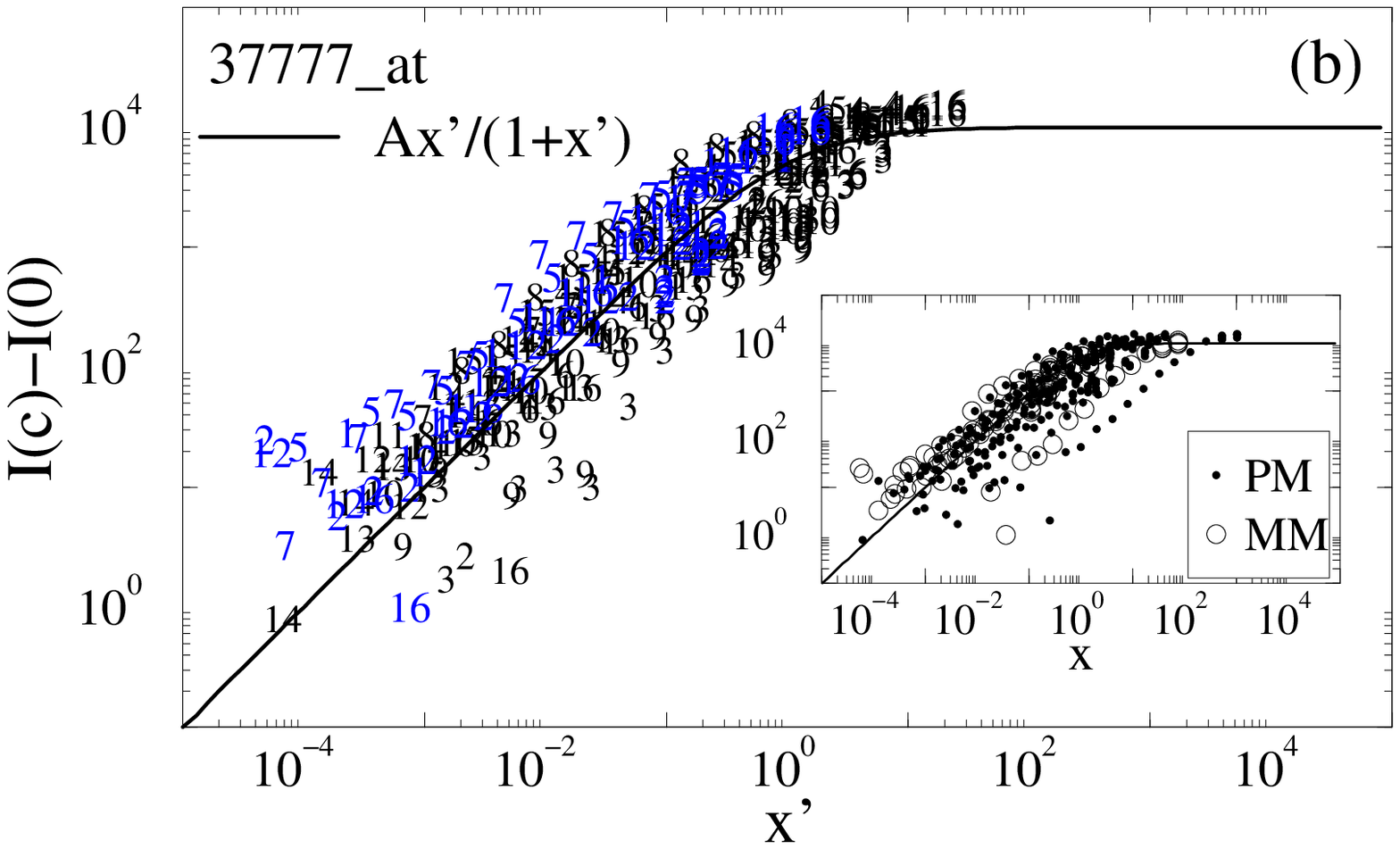}
\caption{Intensities for the probe sets 1024\_at (a) and 37777\_at (b)
plotted as functions of the scaled variable $x'= \alpha c \exp (\beta
\Delta G)$, which takes into account of $\alpha$, the fraction of target
sequences hybridizing in solution.  As a comparison we show in the insets
the same quantity plotted as a function of $x = c \exp (\beta \Delta G)$.}
\label{FIG13}
\end{figure*}

Figure \ref{FIG12} shows the reduction of the effective target
concentration $\alpha_k$ for all the ``spike-in" genes of the experiments
1521 as a function of the free energy of RNA/RNA duplex for each
probe. The parameter $\alpha_k$ is determined from the distance of the
experimental data to the Langmuir isotherm $Ax/(1+x)$ in $\Delta I$ vs
$x$ plots. In Figure \ref{FIG12} the data follow two different behaviors
below and above $45$ kcal/mol. For low hybridization free energies
$\alpha_k$ is constant and roughly equal to 1, indicating that those
regions of the target RNA are single stranded and available for binding
to the probes.  For free energies larger than $45$ kcal/mol $\alpha_k$
diminishes following approximately an exponential decay, due to the
possible effect of enhanced hybridization in solution. We stress that
the free energies shown in Figure \ref{FIG12} are for RNA/RNA duplexes
and these are typically stronger than RNA/DNA or DNA/DNA counterparts.
We fit the global behavior of $\alpha_k$ with the following equation
\be 
\alpha_{\rm k} = \frac 1 {1 + \tilde{c} \exp (-\beta' \Delta G_{\rm R})} 
\label{alpha_k} 
\ee 
where $\Delta G_{\rm R}$ is the RNA/RNA hybridization free energy in
solution. The best fit of the data is shown as a solid line in Figure
\ref{FIG12}, which leads to $\tilde{c} \approx 10^{-2}$ pM and $\beta'
=0.67$ mol/kcal, i.e. $T'= 725^\circ$ K.

The Eq. (\ref{alpha_k}) resembles that for a two state process in which
the target RNA reacts with a fragment with concentration $\tilde{c}$.
In reality there are many different matching fragments hybridizing
with the same target region. One should not view $\tilde{c}$ as a real
concentration, rather the whole $\tilde{c} \exp (-\beta \Delta G_{\rm
R})$ as a global relative probability for hybridization in solution,
which is obtained by averaging over all these processes.

Having now fixed the four fitting parameters $A$, $T$, $\tilde{c}$
and $T'$, we can reanalyze the data collapse by using as a scaling
variable $x' = \alpha_{\rm k} c \exp(\beta \Delta G)$, with $\alpha_k$
given in Eq. (\ref{alpha_k}).  Figure \ref{FIG13} shows the plot of
$\Delta I$ with the new scaling variable for the probe set 1024\_at
(left) and 37777\_at (right). Notice the nice collapse of all PM and MM
intensities into a single master curve, now in much better agreement
with the Langmuir isotherm.  Similar plots for all ``spike-in" genes
show equally good collapses (plots for all 14 genes of the Latin square
set are given in \cite{affy_online}).

\begin{figure}[t]
\includegraphics[width=7cm]{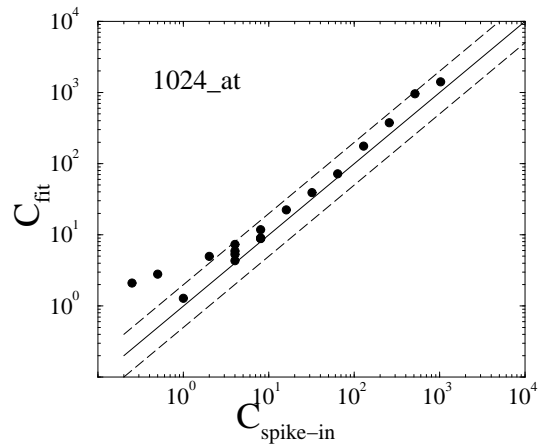}
\caption{Fitted concentration vs. spike-in concentration for the probe set
1024\_at. The solid line is $y=x$, while the dashed lines correspond to
$y=x/2$ and $y=2x$.}
\label{fit}
\end{figure}

We can use the proposed model to fit the experimental data keeping the
absolute target concentration as the only fitting parameter. A plot of
the fitted concentration as a function of the spike-in concentration
is given in the inset of Fig. \ref{fit}. The fitted concentration is,
apart from the region $c_s < 1$ pM, within a factor two from the spike-in
value $c_s$. A result which compares favorably with other algorithms
\cite{heks03_sh,held03,deut04} (for more details see \cite{affy_online}).

\begin{figure*}[t]
\includegraphics[height=5.3cm]{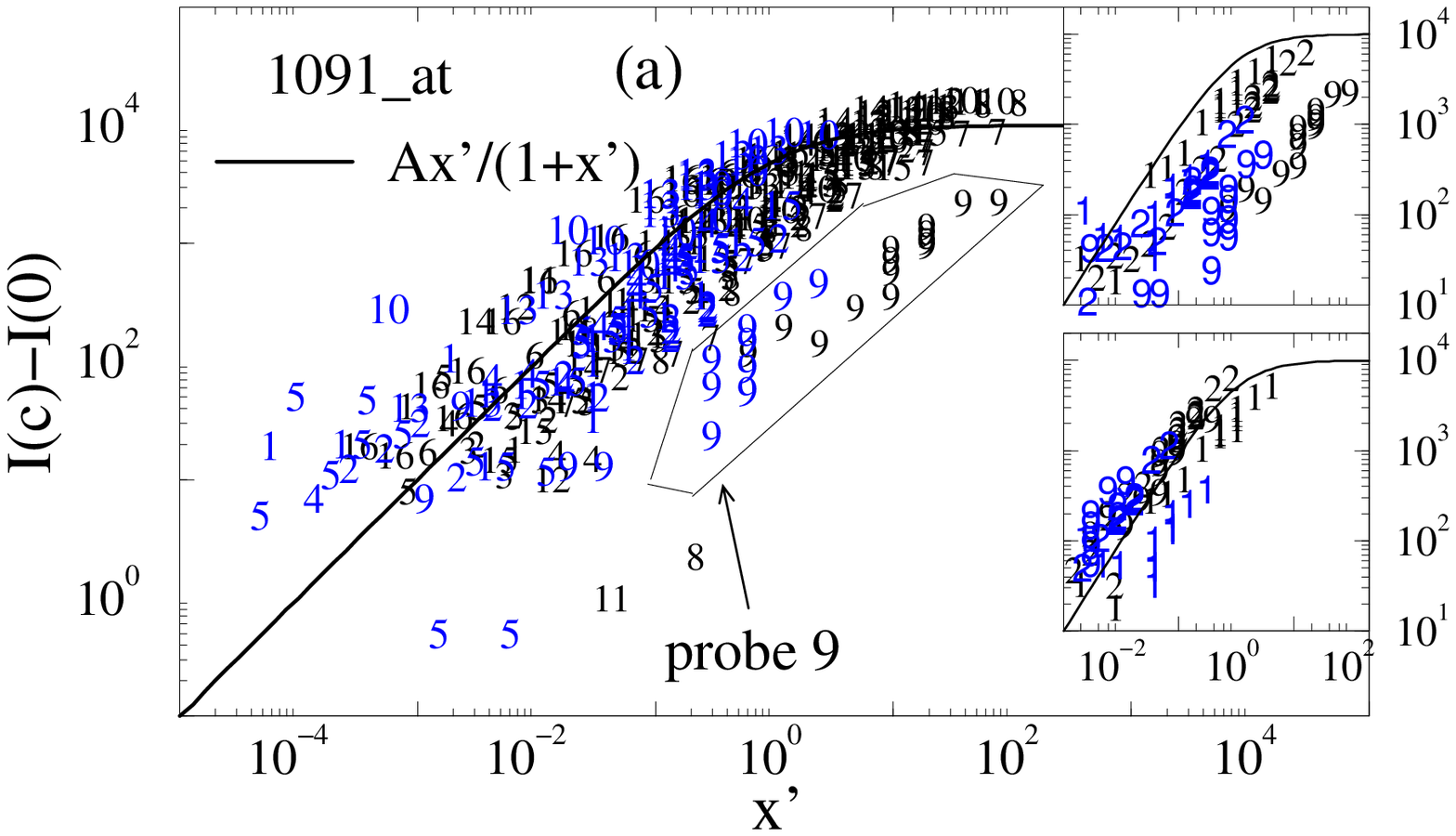}
\ \ \
\includegraphics[height=5cm]{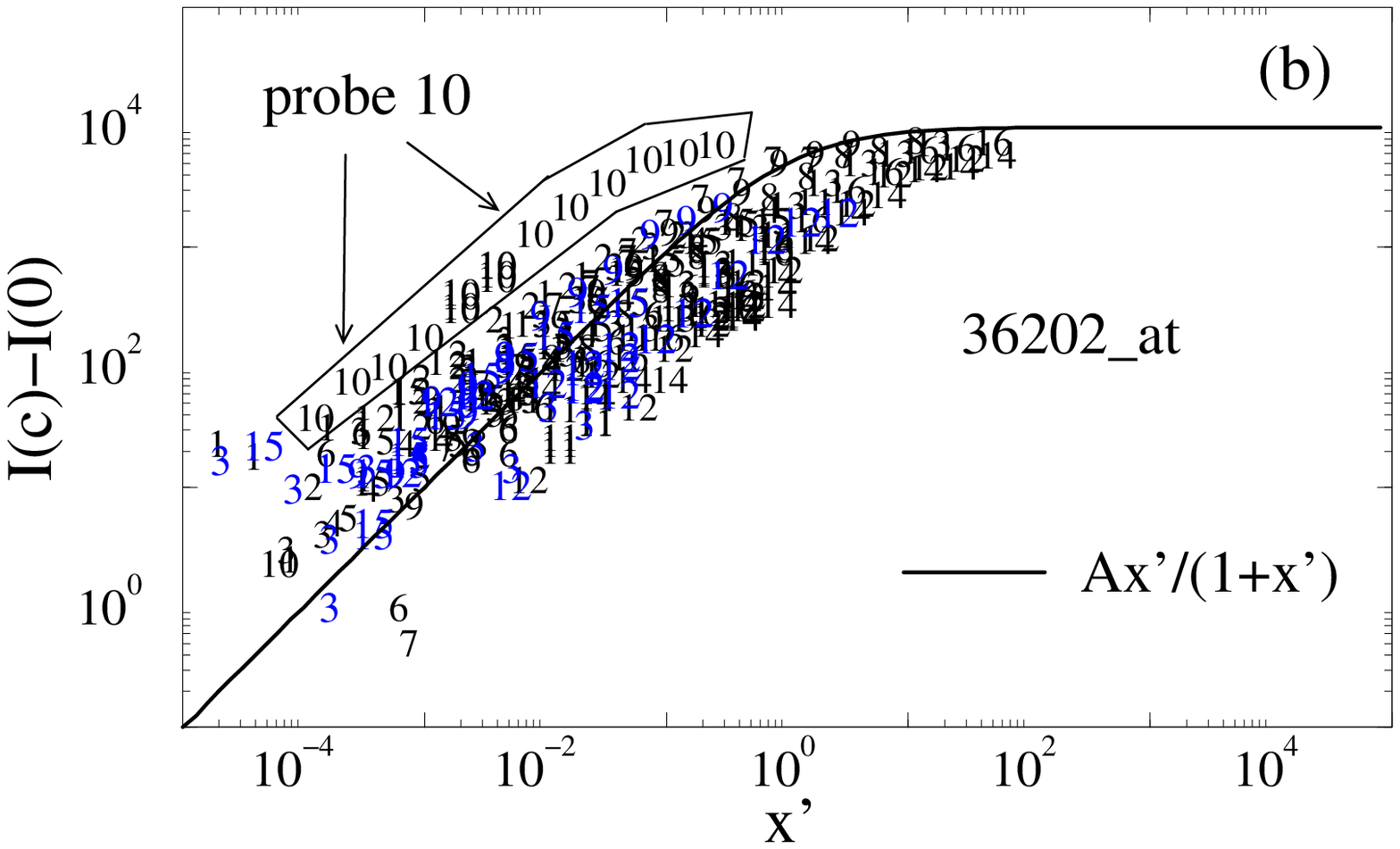}
\caption{Examples of deviations from the rescaled Langmuir isotherm
$Ax'/(1+x')$. (a) The probe 9 of the probe set 1091\_at has substantially
lower signal than that expected. (b) The probe 10 of the probe set
36202\_at has a significantly higher signal than that expected from the
Langmuir isotherm. The insets of (a) show the intensities for the three
``defective" probes, which do not align against the GenBank entries
BC013368.2 and AL833563.1. All these probes have lower intensities
than expected from the Langmuir curve $Ax'/(1+x')$ (upper inset). A
recalculation of the hybridization free energies for these probes leads
to a horizontal shift of the data, which are much  closer to the
Langmuir isotherm (lower inset).}
\label{deviations}
\end{figure*}

\section{Deviations from the modified Langmuir isotherm}
\label{sec:deviations}

An analysis of the 14 spike-in genes reveals that there are still a few
probes deviating from the expected behavior of the modified Langmuir
isotherm $Ax'/(1+x')$, which takes into account the target-target
hybridization in solution. Figure \ref{deviations} shows two examples of
such deviations: (a) the probe 9, both PM and MM, of the probe set 1091\_at
and (b) the probe 10 of the probe set 36202\_at. These deviations are
systematic as they are observed in other replicates of the Latin square
experiments and at all concentrations. Note that the large majority of
the probes are in quite good agreement with the Langmuir isotherm as
in the examples shown in Fig. \ref{FIG13}.  The deviations typically
involve just one probe per probe set and they are observed in very few
of the 14 spike-in genes of the Latin square set.

It is very instructive to look at these deviations more in detail.
We performed a systematic sequence alignment against the whole human
genome stored in public data banks (as GenBank) for all 14 probe sets
of the Latin square experiments. Affymetrix arrays are produced by
photolitographic techniques and each probe is synthesized in situ
using the sequences taken from GenBank. However, the GenBank entries
are continuously updated, and in some sequences errors may be present.
The Affymetrix {\it NetAffx}$^{\rm TM}$ Analysis center \cite{netaffx},
provides information on the GenBank entries used to design the probe
sequences.

Let us discuss first the probe set 1091\_at. {\it NetAffx}$^{\rm TM}$
indicates that this probe set was obtained from the GenBank entry
M65066.1. A sequence alignment indeed shows that all the probe set
sequences for the 1091\_at match fully with the GenBank entry M65066.1.
The alignment also shows that the two other sequences with GenBank
entries BC013368.2 and AL833563.1 match perfectly with 13 of the 16
probes of probe set 1091\_at, while the match is only partial for the
probes 1,2 and 9. The Table \ref{table_blast} summarizes the results
of the alignment for the probe set 1091\_at. The difference is a single
nucleotide close to the $5$' and $3$' ends for the probes 1 and 2, while
there are 5 mismatching nucleotides for the probe 9.  Note also that the
GenBank sequence M65066.1 dates from 1994 (see Table \ref{table_blast}),
while the two other entries are much more recent. We therefore suspect
that the entry M65066.1 contains some annotation errors. As Affymetrix
probe sequences are obtained by public databases, which are constantly
updated, inconsistencies between probes and actual mRNA sequences may
be present in some GenBank entries.  If we assume that BC013368.2 and
AL833563.1 contain the correct mRNA sequence, then the hybridization
free energies that were used in Fig. \ref{deviations}(a) for the probes
1, 2 and 9 are overestimated and need to be corrected. Note that the
three probes 1,2 and 9 have all intensities lower than expected from
the Langmuir model as shown in the upper inset of Fig. \ref{FIG13}(a).

\begin{table*}[t] 
\caption{
Best alignments for the probes 1, 2 and 9 of the probe set
1091\_at. For each probe the first line is the sequence found in the
Affymetrix chip, which aligns perfectly with the sequence with GenBank
entry M65066.1 (second line).  The sequences with GenBank entries
BC013368.2 and AL833563.1 align perfectly with 13 of the probes in the
probe set 1091\_at, but they have some differences with the probes 1,
2 and 9. The differing nucleotides are underlined.  The last column of
the Table shows the sequence submission date to the GenBank.}
\vspace{5mm}
{\footnotesize
\begin{tabular}{@{}cclcc} 
Probe   & Origin        & Sequence &  GenBank & Date \\ 
\hline
	&  Affymetrix	&$5'$-TATGAGATTGATCTTGCCCCTAATT-$3'$&		  & \\
   1	&  Blast 1	&$5'$-TATGAGATTGATCTTGCCCCTAATT-$3'$&  M65066.1 & 10-NOV-1994 \\
	&  Blast 2	&$5'$-T{\underline G}TGAGATTGATCTTGCCCCTAATT-$3'$& BC013368.2 &
19-NOV-2003\\
	&  Blast 3	&$5'$-T{\underline G}TGAGATTGATCTTGCCCCTAATT-$3'$& AL833563.1 &
13-MAY-2003\\
\hline
	&  Affymetrix	&$5'$-GCAGAAGTCAAGCCAGCCGCGGCCC-$3'$& &\\
   2	&  Blast 1	&$5'$-GCAGAAGTCAAGCCAGCCGCGGCCC-$3'$&  M65066.1 & 10-NOV-1994 \\
	&  Blast 2	&$5'$-GCAGAAGTCAAGCCAGCCGCGG{\underline G}CC-$3'$&  BC013368.2 &
19-NOV-2003\\ 
	&  Blast 3	&$5'$-GCAGAAGTCAAGCCAGCCGCGG{\underline G}CC-$3'$&  AL833563.1 & 
13-MAY-2003\\ 
\hline
	&  Affymetrix	&$5'$-CTGTCCTTGGTCCG\_\_CATGGCTCGTT-$3'$& &\\
   9	&  Blast 1	&$5'$-CTGTCCTTGGTCCG\_\_CATGGCTCGTT-$3'$&  M65066.1 & 10-NOV-1994 \\
	&  Blast 2	&$5'$-CTGTCCTTGGTCCG\underline{AGGCT}GCTCGTT-$3'$& BC013368.2&
19-NOV-2003\\ 
	&  Blast 3	&$5'$-CTGTCCTTGGTCCG\underline{AGGCT}GCTCGTT-$3'$& AL833563.1&
13-MAY-2003\\
\end{tabular} 
}
\label{table_blast} 
\end{table*}

Before discussing the free energies corrections, we recall that
the Affymetrix RNA target in solution is actually anti-sense RNA,
complementary to the usual mRNA sequences. Therefore the surface-bound
probes have the same sequences as mRNA's, apart from the substitution
of U with T.

{\it Probe 1}: The new mRNA annotation from the sequences BC013368.2
and AL833563.1 of Table \ref{table_blast} implies that a CA mismatch
with a triplet rA{\underline C}A/dT{\underline A}T is formed when
the target RNA hybridizes with the probe 1. There is no information
in the present literature  \cite{sugi00} about this mismatch triplet,
therefore we cannot assign a free energy to it. We notice anyhow that the
mismatch occurs very close to the $5'$ end of the probe sequence (see Table
\ref{table_blast}), which is the ``free" end as the probes are linked to
the substrate at their $3'$ end. It is plausible that close to its free end
the double helix RNA/DNA can be substantially distorted without a large
penalty in free energy. The bases in the mismatch rC/dA should be still
able to form two hydrogen bonds, therefore we consider likely that this
mismatch does not affect substantially the hybridization free energy.
This is a plausible explanation of the fact that the probe 1 (both PM
and MM) deviates only slightly from the Langmuir isotherm, as shown in the
inset of Fig. \ref{deviations}(a).

{\it Probe 2}: The new annotation implies an extra mismatch of the type
rG{\underline C}C/dC{\underline C}G, for which no free energy has been
given in the literature. The free energy differences between perfect
triplets and triplets with a CC mismatch, as given in Figure \ref{FIG14},
suggest as a rough estimate for the CC mismatch of about $4.5$ kcal/mol.
This causes a shift of the data toward a lower value of the variable $x'$
and shifts them much closer to the curve $Ax'/(1+x')$.

\begin{figure}[t]
\includegraphics[width=8cm]{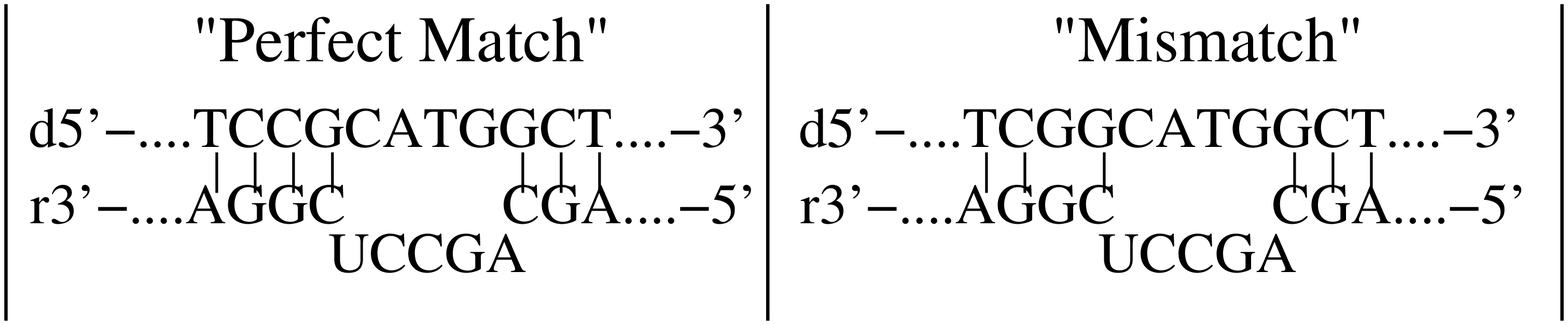}
\caption{Expected duplexes formed with the ``defective" probe 9 of the set
1091\_at for the PM and MM signal, if the mRNA sequence is taken from the 
entries BC013368.2 and AL833563.1. The upper strand is the surface-bound DNA, 
while the lower one is the RNA target.}
\label{FIG06}
\end{figure}

{\it Probe 9}: For the probe 9 the alignment of the sequence M65066.1 with
those in BC013368.2 and AL833563.1 differs the most. The duplexes formed
in this case are as shown in Figure \ref{FIG06} and contain an inner
asymmetric loop with two arms with 4 and 5 nucleotides. It is difficult
to evaluate the hybridization free energies for these configurations. A
rough estimate, taking $\Delta G^{\rm loop} = 0$, yields a free energy
shift of $8-10$ kcal/mol.


We have used the estimated free energies to correct for the $x'$ variables
for the probes 2 and 9. For instance a shift of $8$ kcal/mol for the
probe 9 implies a correction factor of $\exp (-8/RT) \approx 3 \cdot
10^{-3}$, where we have used $T=700^\circ$ K. Analogously for the probe
2 we find $\exp (-4/RT) \approx 0.05$. The insets in Fig. \ref{FIG06}(a)
show the intensities for the three probes with conflicting alignments
in the case of un-normalized data (top) and data with rescaled factors
for the probes 2 and 9 (bottom). In the latter case the agreement with
the Langmuir isotherm is substantially improved.

In order to assess on the possible frequency of annotation errors we have
performed an alignment analysis of all probe sets of the Latin square
set (for more details see \cite{affy_online}).  The only potential
annotation problems were those detected for the probe set 1091\_at,
which suggests that these errors should not be too frequent, at least
in the human genome.

We turn now to the probe 10 of Fig. \ref{deviations}(b), which has a
signal significantly higher compared to the Langmuir prediction. In the
whole set of Latin square data we found only another example of a similar
high signal, namely the probe 16 of the probe set 36085\_at.  Analyzing
these two sequences we find that they share a common feature: both are
A-rich close to the 3$'$ end. The sequences are \ldots CACAAAAG-3$'$
(36202\_at10) and \ldots CAATAAA-5$'$ (36085\_at16). Note that the Table
\ref{tableI} gives for the combination rUU/dAA the lowest free energy.
A possible improvement of the data collapse could be obtained by
introducing position-dependent weights $w_i$ with $i=1$, $2$ \ldots
so that
\be
\Delta G = \Delta G^{\rm init.} + \sum_{i=1}^{24} w_i \Delta G_i
\label{wi}
\ee
where $\Delta G_i$ is taken from Table \ref{tableI} and the sum is
extended over the nucleotides of the target sequence. One could assume
that far from the substrate $w_i \approx 1$, so that the hybridization
free energy reaches the solution limit and that $w_i < 1$ close to the
substrate. A reweighting of this type lowers the global value of $\Delta
G$ (and consequently decreases the effective fitting temperature) and
could lower the anomalously high signal of the two probes A-rich at
their 3$'$ end. It is not clear a priori which function to choose for
the weight $w_i$, apart from the fact that it should increase far from
the surface, and different possibilities will be explored elsewhere.
We stress however that the choice $w_i = 1$, used in this paper, provides
already a quite good fit of the experimental data except in very few cases.

\section{Discussion}
\label{sec:Conclusion}

In this paper we have analyzed a series of controlled experiments on
Affymetrix microarrays using a simple model of RNA/DNA hybridization. We
have shown that for each probe (PM and MM) of a probe set the intensities
plotted as a function of the free energy of RNA/DNA hybridization
tend to align along a single master curve in quite good agreement
with a Langmuir isotherm.  In fact the intensities of half of the
``spike-in" genes are already well fitted with the simple form given in
Eq. (\ref{langmuir}). For the other half, those containing probes with
higher CG content, we found that one has to include the effect of target
hybridization in solution, which diminishes the effective concentration
of single stranded RNA sequences in solution. This effect is well
described by a simple analytical form given by Eq. (\ref{alpha_k}).
Despite this very simplified model the fit with the experimental data
is very satisfactory, as shown by the scaling collapses (i.e. plots of
intensities as function of the rescaled variable $x'$ of Eq. (\ref{x_k})).
Although the data are somewhat noisy the calculations of the target
concentrations are in good agreement with the input spike-in values
(see Fig. \ref{fit} and \cite{affy_online}). This is due to the fact
that concentrations are obtained from averaging over the signal of each
individual 16 PM probes and of the MM probes of which we were able to
include in the analysis. The averaging over these data points yields
quite robust and reliable estimated of target concentration values.

The feature that is still missing in our analysis is the calculation
of the background level ($N$ in Eq. (\ref{specific_non-specific})). We
circumvented this problem by subtracting from the intensities those
measured at zero spike-in concentration. This is only possible for the
Latin square data. A good estimate of the background level could help
in improving the quality of the fits in the low concentrations limit.
These issues will be considered elsewhere.

The physics of the hybridization in high density microarrays has been
investigated recently by other groups. We comment now on the differences
between the present approach and what has been done in the literature
so far.  In a recent paper Hekstra {\it et al.} \cite{heks03_sh} found
nice agreement of the Latin square data with a Langmuir model. Their plots
of rescaled intensities versus rescaled concentrations follow very well
the curve $x/(1+x)$, with small fluctuations. For their rescaling they
use probe-dependent values, a procedure which requires the use of $24$
fitting parameters. The advantage of our approach is that we find good
collapses of the experimental data using a very simple model with only
few fitting parameters.


It has been recently claimed that the MM probes do not follow the behavior
predicted by the standard hybridization theory \cite{naef02}. Our
analysis, instead, shows that MM probes intensities follow the same
Langmuir isotherm as the PM probes. For the mismatches we used
the trinucleotide free energies for RNA/DNA duplexes in solution
\cite{sugi00}. A very important aspect of the MM hybridization,
as highlighted in studies of RNA/DNA duplexes melting in solution
\cite{sugi00}, is that their free energy strongly depends on the type
and order of the two nucleotides close to the mismatch. This is probably
the reason why an analysis of the mismatches based on single base pairs
energies, as in Ref. \cite{naef03}, shows deviations from the Langmuir
isotherm of the PM probes. We believe that the strong dependence of the
mismatch free energy on the two neighboring nucleotide is a very important
aspect for the correct modeling of the hybridization of MM probes.

In Ref. \cite{held03}, the Langmuir model for target-probe hybridization
was used to fit Affymetrix Latin square data. The hybridization free
energies were obtained from values of DNA/DNA duplexes in solution
\cite{sant98}, and not for RNA/DNA duplexes, as we have done here.
As discussed in Section \ref{sec:RNA_DNA} there are some differences
between the two sets of parameters. An explicit example emphasizing the
influence of these differences for the fitting procedure is shown in
the Appendix \ref{sec:app_held}.  Our results show that a correct free
energy parametrization improves substantially the quality and stability
of the fits.

In other studies \cite{zhan03,naef03}, the free energies were obtained
directly from a fit of Affymetrix data assuming a input relationship
$I(\Delta G)$.
The binding free energies were taken dependent on the distance from the
substrate \cite{zhan03,naef03}, while we have so far calculated free
energies by summing up uniformly over all the stacking energies of the
probe sequences.  As pointed out before, a position dependent weight in
the free energy calculation may improve the quality of our data collapses
for those few A-rich probes close to the substrate which we found to
deviate more strongly from the Langmuir model.  The overall quality of
the fits remains however quite good also in absence of position-dependent
binding (see Ref. \cite{affy_online}).

Another effect which has been claimed to be relevant for hybridization
in high density DNA microarrays is the Coulomb interaction between
a highly negatively charged surface DNA layer and negatively charged
target molecules \cite{vain02,halp04}. These effects may play a role
for a system with monodisperse probe length distribution. However, in
Affymetrix chips the probe lengths are widely distributed \cite{form98},
An analysis of the electrostatic interaction \cite{affy_online}, shows
that its strength is much weaker compared to that  of the systems
studied in Refs. \cite{vain02,halp04}. It is thus possible to neglect
electrostatic effects, as we did here and as done in other studies
involved with the physical modeling of hybridization in Affymetrix arrays
\cite{zhan03,naef03,heks03_sh,held03}.


Finally, one may wonder how representative the spike-in targets chosen
by Affymetrix are for the overall behavior of the microarray.  A recent
investigation \cite{leo} of several human housekeeping genes (i.e. those
which are expressed in virtually all tissues) and of the Affymetrix
spike-in data for the chipset HGU133 shows that the intensities within
a given probe set follow a distribution which is very similar to that
observed in this work.

\acknowledgements

E.C. would like to thank the Isaac Newton Institute for Mathematical
Sciences in Cambridge, where this work was started, for kind hospitality.
We are grateful to G. Barkema and J. Klein-Wolterink for interesting
discussions. We acknowledge financial support from the Van Gogh Programme
d'Actions Int\'gr\'ees (PAI) 08505PB of the French Ministry of Foreign
Affairs.

\begin{table}[t]
\caption{
$\Delta G_{\rm 37}$ for triplet mismatches in RNA/DNA duplexes, 
where the upper strand is RNA (the orientation is $3'$ to $5'$ from left to right)
and lower strand DNA. Only the mismatches which are realized in the Affymetrix 
chip are shown.
{\it a}: Deduced from the rUdG mismatches, 
{\it b}: Deduced from the rAdA mismatch.
{\it c}: Deduced from the rCdT mismatches.
{\it d}: Deduced from the rCdA mismatches.
Between parenthesis is the free energy of a perfectly matching triplet obtained by
interchanging C with G and A with T in the central nucleotide of the DNA strand.}
\vspace{5mm}
\begin{tabular}{cc|cc}
Sequence & $\Delta G_{\rm 37}$(kcal/mol)  & Sequence & $\Delta G_{\rm 37}$(kcal/mol)\\
\hline
GG-mismatches &&&\\
${\rm rC}{\bf G}{\rm G} \atop {\rm dG}{\bf G}{\rm C}$ &  0.11 \ (-4.6)&
${\rm rC}{\bf G}{\rm C} \atop {\rm dG}{\bf G}{\rm G}$ & -0.97 \ (-4.4)\\
&&&\\
${\rm rG}{\bf G}{\rm G} \atop {\rm dC}{\bf G}{\rm C}$ & -1.26 \ (-5.8)&
${\rm rG}{\bf G}{\rm C} \atop {\rm dC}{\bf G}{\rm G}$ & -2.25 \ (-5.6)\\
&&&\\
${\rm rA}{\bf G}{\rm A} \atop {\rm dT}{\bf G}{\rm T}$ &  0.48$^a$ \ (-3.1)&
${\rm rA}{\bf G}{\rm C} \atop {\rm dT}{\bf G}{\rm G}$ & -0.62$^a$ \ (-4.5)\\
&&&\\
${\rm rC}{\bf G}{\rm U} \atop {\rm dG}{\bf G}{\rm A}$ & 1.24$^a$ \ (-2.8)&
${\rm rU}{\bf G}{\rm G} \atop {\rm dA}{\bf G}{\rm C}$ & 0.67$^a$ \ (-4.5)\\
&&&\\
\hline
AA-mismatches &&&\\
${\rm rC}{\bf A}{\rm G} \atop {\rm dG}{\bf A}{\rm C}$ & 1.05 \ (-2.7)&
${\rm rC}{\bf A}{\rm C} \atop {\rm dG}{\bf A}{\rm G}$ & 0.32 \ (-3.0)\\
&&&\\
${\rm rG}{\bf A}{\rm G} \atop {\rm dC}{\bf A}{\rm C}$ & 0.20 \ (-3.1)&
${\rm rG}{\bf A}{\rm C} \atop {\rm dC}{\bf A}{\rm G}$ & 0.29 \ (-3.4)\\
&&&\\
\hline
UT-mismatches &&&\\
${\rm rC}{\bf U}{\rm G} \atop {\rm dG}{\bf T}{\rm C}$ & 1.05$^b$ \ (-2.5)&
${\rm rC}{\bf U}{\rm C} \atop {\rm dG}{\bf T}{\rm G}$ &   0.25   \ (-2.4)\\
&&&\\
${\rm rG}{\bf U}{\rm G} \atop {\rm dC}{\bf T}{\rm C}$ &  0.44 \ (-2.7)&
${\rm rG}{\bf U}{\rm C} \atop {\rm dC}{\bf T}{\rm G}$ & -0.22 \ (-2.6)\\
&&&\\
\hline
CC-mismatches &&&\\
${\rm rC}{\bf C}{\rm G} \atop {\rm dG}{\bf C}{\rm C}$ & 0.73$^c$ \ (-3.8)&
${\rm rC}{\bf C}{\rm C} \atop {\rm dG}{\bf C}{\rm G}$ &   -      \ (-4.2)\\
&&&\\
${\rm rG}{\bf C}{\rm G} \atop {\rm dC}{\bf C}{\rm C}$ & 0.28$^d$ \ (-4.4)&
${\rm rG}{\bf C}{\rm C} \atop {\rm dC}{\bf C}{\rm G}$ &   -      \ (-4.8)\\
&&&
\end{tabular}
\label{tableII}
\end{table}

\appendix
\section{Free energies for mismatches}
\label{sec:appendix}

The Table \ref{tableII} shows all the free energies for triplets with
a single mismatch used in this paper. Part of these free energies are
obtained from experimental results of Ref. \cite{sugi00}. In some cases
the free energies were deduced from the analogy with other mismatches.
For instance, as pointed out in Ref. \cite{sugi00}, the free energies
for triplets with rAdA mismatches are close to those of triplets with
rUdT mismatches. In absence of experimental determinations of mismatch
free energies for the triplet rC\underline{U}G/dG\underline{T}C,
we assign to the latter the same free energy as the mismatch
rC\underline{A}G/dG\underline{A}C, which is of $1.05$ kcal/mol.  Of the 18
triplet free energies in Table \ref{tableII}, 11 were obtained by direct
experimental data inputs, while 7 from similarities with other mismatches.

\begin{figure}[t]
\includegraphics[width=8cm]{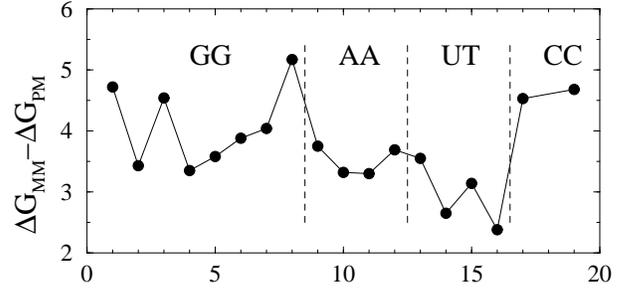}
\caption{Free energy differences between MM and PM triplets for the 18
mismatches given in Table \ref{tableII}, where the labeling follows
the same ordering as the Table. As an example the first point is the
free energy difference between rC{\underline G}G/dG{\underline G}C
and rCGG/dGCC, the second point the free energy difference between
rC{\underline G}C/dG{\underline G}G and rCGC/dGCG \ldots}
\label{FIG14} 
\end{figure} 

Another interesting quantity is the free energy difference between a
perfect matching triplet and one with a central mismatch, i.e. the free
energy shift due to a single mismatch.  Figure \ref{FIG14} reports
the free energy differences for the 18 mismatches  given in Table
\ref{tableII}, following the same order. These differences are obtained
by subtracting from the data in Table \ref{tableII} the values between
parenthesis, corresponding to a perfect matching triplet. The Figure
\ref{FIG14} shows that the free energy difference is quite sensitive to
the type of mismatch and of its two neighboring nucleotides.

\section{Comparing RNA/DNA with DNA/DNA hybridization free energies}
\label{sec:app_held}

\begin{figure}[t]
\includegraphics[width=8cm]{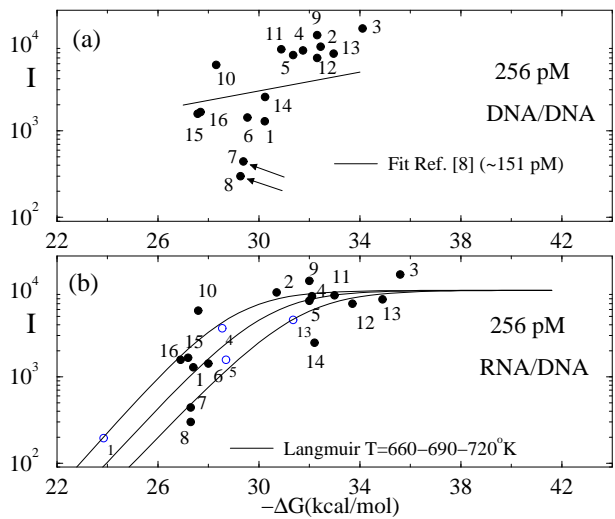}
\caption{Intensity for the probe set 36311\_at for a target
concentration of $256$ pM plotted as function of (a) DNA/DNA and
(b) RNA/DNA hybridization free energies. (a) reproduces Figure 5
of Ref. \cite{held03}, and the solid line is the best fit with the
parameters used in that reference; the probes 7 and 8 (indicated as
arrows) are considered as outliers.  In the case (b) the solid lines
are the Langmuir isotherms from Eq. (\ref{langmuir}), $A=10^4$, $c=256$
pM and three values of the temperature.}
\label{FIG15}
\end{figure}

The approach followed in Ref. \cite{held03} uses hybridization
free energy for DNA/DNA duplexes in solution, while throughout this
paper we used RNA/DNA parameters, as in the Affymetrix Latin square
experiments the target is composed by RNA, while the probes are
surface-bound DNA sequences.  In order to illustrate the difference in
the intensity vs. free energy plots we show in Figure \ref{FIG15} the
fit to the Langmuir model the intensities of the probe set 36311\_at
(from the Affymetrix file 1521g99hpp\_av06) when (a) DNA/DNA and (b)
RNA/DNA hybridization free energies are used. The figure shows the
same data of Figure 5 of Ref. \cite{held03}. The best fit obtained
from the parameters of Ref. \cite{held03} is the thick solid line
shown in Figure \ref{FIG15}(a), where the probes 7 and 8, indicated
by arrows, were considered as outliers. In Figure \ref{FIG15}(b) the
solid lines are the Langmuir isotherms with the same parameters as
in Figures \ref{FIG03}(a) and \ref{FIG04}(a).  By comparing the two
plots one can conclude that the fit in Figure \ref{FIG15}(b), which is
also consistent with plots for other probe sets (see Figs. \ref{FIG03}
and \ref{FIG04}), is more convincing than that shown in the case (a).
Note that in Fig. \ref{FIG15}(b) the four MM intensities follow the
same behavior as the PM probes.  The analysis of Ref. \cite{held03}
is restricted to PM probes only.

The difference between the fits performed here and those reported
in Ref. \cite{held03} are also due to a different approach to the
analysis.  In Ref. \cite{held03} the intensities for each probe are
fitted as a function of the concentration $c$ using the three adjustable
parameters $A$, $K= \exp(- \beta \Delta G)$ and the background level $N$.
Although a three parameters fit appeared to reproduce experimental data
for different probes very well \cite{held03} one of the problems with
this analysis is that $A$ was found to vary over one order of magnitude
from probe to probe. As this is an unphysical feature, $A$ was then kept
constant for all probes while the other two parameters were allowed to
vary \cite{held03}. Ref. \cite{held03} reports as best fitted value $A
\approx 9\ 500$, which compares favorably to our choice $A=10^4$. The
problem is that the fits of Intensities vs. concentration obtained by
binning over different free energies are less convincing (see Fig. 3 of
Ref. \cite{held03}). Analyzing then the decay of $1/K$ as function of the
DNA/DNA hybridization free energy an effective temperature of $T \approx
2\ 100^\circ$ K is found. This is roughly three times higher of what
we find in this paper, from a direct analysis of plots of intensities
vs. RNA/DNA hybridization free energies.  The use of an accurate free
energy parametrization is very important: as $\Delta G$ is related to the
intensity thorough an exponential factor, small variations of $\Delta G$
estimates may have profound influences on the values and robustness of
the fitting parameters.


\end{document}